\documentstyle[aps,prb,twocolumn,epsfig]{revtex}
\begin{document}

\bibliographystyle{../prsty}
\title{A dynamic scattering approach for a gated interacting wire }
\author{In\`es Safi}
\address{Service de Physique de l'\'Etat Condens\'e, Centre d'\'Etudes de Saclay\\ 91191 Gif-sur-Yvette, France\\}

\maketitle

\begin{abstract}
A new scattering approach for correlated one-dimensional systems is
developed. The adiabatic contact to charge reservoirs is encoded in
time-dependent boundary conditions. The conductance matrix for an arbitrary
gated wire, respecting charge conservation, is expressed through a
dynamic scattering matrix. It is shown that the dc conductance is equal to $%
e^2/h$ for any model with conserved total left- and right-moving charges.
The ac conductance matrix is explicitly computated for the interacting
Tomonaga-Luttinger model (TLL).
\end{abstract}

Pioneered by Landauer,\cite{scattering} the scattering approach for quantum
transport has proven powerful in mesoscopic physics. Nevertheless, it is
restricted to non-interacting systems, and to the stationary regime. There
were formal extensions to finite frequency transport based on a
self-consistent approach,\cite{butticker_pretre} or non-equilibrium
techniques for interacting dots,\cite{AC_keldysh} but these formalisms are difficult
to exploit. Proposed here is a new scattering approach at arbitrary
frequency for linear transport through a strongly correlated, one-dimensional
wire in the low-energy regime. Charge reservoirs connected adiabatically to the
wire are accounted for by appropriate boundary conditions. Coupling
to a gate is taken into account, ensuring charge conservation. The
corresponding AC $3\times 3$ conductance matrix is expressed through a novel
dynamic ``scattering'' matrix ${\bf S}(\omega )$. Further progress is then
made in two cases. First, for any model where the total charge for right-
and left-moving electrons is conserved, the transmission is shown to be
 unity in the zero-frequency limit. This generalizes the DC conductance
result $g=e^2/h$ shown for a Tomonaga-Luttinger liquid (TLL)\cite{ines,maslov_pon,DC} or arbitrary
finite-range interactions. \cite{ines_bis} The same result was shown in Ref.%
[\onlinecite{alekseev_goldstone}] through different arguments restricted to the
stationary regime, without describing the reservoir-wire interface.
Second, ${\bf S}(\omega )$ is computed for the TLL model, giving an AC
conductance that depends on interactions in contrast to the stationary regime.

Without connecting one-dimensional leads to an interacting wire, this work extends the concept introduced in Refs.
[\onlinecite{ines,ines_proc,these_annales}] where reservoirs are simulated by
the electrons they inject. The leads have served to define the incident and transmitted 
electrons, different from the proper modes of the wire. ${\bf S}(\omega )$
will be related to the dynamic transmission. For a TLL model, the
same conductance results were found by the author by computing the current
in response to an appropriate external electric field.\cite{ines_proc,these_annales}
 More recently, they were confirmed by Blanter et al\cite{blanter} through a self-consistent
treatment of interactions, justified in the absence of backscattering. Other works based on the Kubo formula in a TLL with leads found different results due to a different
electric field profile to which current is very sensitive.\cite{AC_Kubo} 

 An
underlying hypothesis of Landauer's approach for noninteracting systems
\cite{scattering} is the ideal nature of the contacts, ensuring that emerging electrons
are absorbed without reflection by the reservoirs.\cite{scattering} Such a
concept cannot be extended to interacting systems, as illustrated in Ref.%
[\onlinecite{ines}].\cite{kane_fisher_contacts,remark_chamon} Rather, interactions give rise
to collective excitations, or Laughlin quasiparticles in edge states, that
are different from the electrons in the reservoirs. An emerging
``quasiparticle'' undergoes a quasi-Andreev type reflection \cite{ines,ines_proc,these_annales,sandler} at a perfect contact with a
reservoir.

 This paper is mainly concerned with systems connected
locally to reservoirs, such as quantum wires, or nanotubes; edge states
couple differently to reservoirs,\cite{remark_chamon} and the present
scattering approach will be extended elsewhere.\cite{ines_hall}

Consider an arbitrary one-dimensional finite wire delimited by $[-a,a].$ The
long wavelength part of the electronic density can be decomposed into right and
left-moving electron densities\cite{bosonisation} $\rho _{+}$ and $\rho _{-}$ including
implicitly the zero modes, $\rho =\rho _{+}+\rho _{-}$
where spin is ignored for simplicity. For $r=\pm $, the boson field $\Phi _r$
defined by $\rho _r=-\partial _x\Phi _r/2\pi $ is the canonical conjugate to 
$r\rho _r$ (Kac-Moody algebra). The kinetic Hamiltonian is $%
H_{kin}=\int_{-a}^ahv_F(\rho _++\rho_-^2)/2$. Any interaction
Hamiltonian $H_{int}$ either between electrons or with impurities can be
expressed as a functional of $\Phi _{+},\Phi _{-}$, thus the total
Hamiltonian, 
\begin{equation}\label{H}
H=H_{kin}+H_{int}+eV_{gate}Q=H(\rho _{+},\rho _{-}),
\end{equation} is a functional of $\rho _{\pm }$. Coupling to a gate is incorporated, and $%
Q=\int_{-a}^a\rho (x)$. 

The current field $j(x)$ can be expressed independently of the dynamics, in or out-of-equilibrium. For this, the Hamiltonian $H^{(A)}$ in the presence of a vector potential $A$, is used :
\begin{equation}
\label{ja}j(x)=-\left. \frac{\partial H^{(A)}}{\partial A(x)}\right| _{A=0}.
\end{equation}
 $A$ can be absorbed by a gauge
transformation of the right and left-going Fermion fields, $\Psi _r\sim
e^{ir\Phi _r}$ for $r=\pm $.\cite{bosonisation} This is accomplished by the substitution\cite
{alekseev_goldstone}%
$$
\Phi _r(x)\rightarrow \Phi _r(x)-\frac{re}\hbar \int^xA(x^{\prime
})dx^{\prime }. 
$$
 Taking the spatial derivative, one obtains $H^{(A)}$ as 
$$
H^{(A)}=H\left( \rho _{+}+\frac eh A,\rho _{-}-\frac eh A\right) . 
$$
Differentiating with respect to $A$ [Eq. (\ref{ja})] yields\cite{remark_oreg}
\begin{equation}
\label{jmu}j(x)=\frac eh\left[ \mu _{+}(x)-\mu _{-}(x)\right].
\end{equation}
For $r=\pm$, $\mu_r$ are operators that play a central role :
\begin{equation}
\label{mur}\mu _r(x)=\frac{\partial H}{\partial \rho _r(x)}=hv_F\rho _r+
\frac{\partial H_{int}}{\partial \rho _r(x)}+eV_{gate}.
\end{equation}
Also of use will be their sum,
\begin{equation}\label{mu}
\mu(x)=\frac{\partial H}{\partial \rho (x)}=\frac 12\left[\mu _{+}(x)+\mu _{-}(x)\right]. 
\end{equation}
\begin{figure}
\begin{center}
\epsfig{file=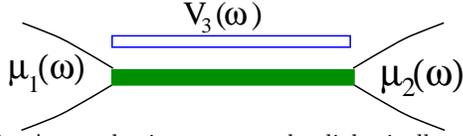,width=6cm}
\caption{A gated wire connected adiabatically to charge reservoirs with time-dependent electrochemical potential $\mu_{1,2}(t)=e^{i\omega t }\mu_{1,2}(\omega)$. The boundary conditions apply in the presence of arbitrary backscattering inside the wire.}
\end{center}
\label{f:gatedw}
\end{figure}
Consider now a typical transport measurement, where one connects the wire
adiabatically at $\pm a$ to charge reservoirs [see Figure].
Spatial and temporal structure on the scale of $\lambda _F$ (Friedel
oscillations) are ignored. The local Hamiltonian density ${\mathcal H}(\pm a)$ is
quadratic in $\rho _{\pm }$ when backscattering is irrelevant, so that $\mu
_{\pm }$ $(\pm a)$ are linear in $\rho _{\pm }$. One can think of the
expectation value of $\mu _{+}(x)$ as the energy required to add a
right-going electron at $x.$ It includes the local Fermi energy $hv_F\rho
_{+}(x)$, the interaction energy, and the gate potential that shifts the
bottom band. In some sense, it is a local electro-chemical potential for
right-going electrons. On the other hand, the left (right) reservoir injects
bare right (left)-going electrons with a well defined electro-chemical
potential $\mu _1$ ($\mu _2$). At the contacts where both incident fluxes
impinge, the energy is conserved in the absence of any dissipation
processes. Thus the field $\mu _{+}(-a,t)$ [$\mu _{-}(a,t)$] is required to
be pinned to $\mu _1$ [$\mu _2$] at any time. To extend this conjecture to
alternative regimes, time variation has to be slow enough so that the reservoirs are driven adiabatically through a sequence of equilibrium
states with well defined time-dependent electro-chemical potential: 
\begin{eqnarray} \label{conditions}
\mu_+(-a,t)&=&\mu_1(t)\nonumber\\
\mu_-(a,t)&=&\mu_2(t).
\end{eqnarray}
Without interactions near the contacts, $\mu _{\pm }(\mp a,t)=hv_F\rho _{\pm
}(\mp a,t)+eV_{gate}(t)$; Eq. (\ref{conditions}) imposes the density for
incident electrons, and generalizes the Landauer concept to AC transport with arbitrary backscattering. But in the presence of interactions, $\mu
_{\pm }$ depend on both $\rho _{+}$ and $\rho _{-}$, thus the density for
incident electrons is not imposed, in contradiction with Ref.%
[\onlinecite{egger}]. Rather, electrons are partially reflected, giving rise
to a contact resistance. Indeed, Eq. (\ref{conditions}) leads to a
discontinuous local electrostatic potential $V_{loc}$. $eV_{loc}$ follows the
electro-chemical potential on the reservoir side, while
\begin{equation}
\label{Vloc}eV_{loc}(x)=\frac{\partial [H-H_{kin}]}{\partial \rho (x)}=\mu(x)
-\frac{hv_F}2\rho(x)
\end{equation}
on the interacting side.\cite{these_annales,ines_bis}Note that $eV_{gate}$ is included. $\mu(x)$ is a local electro-chemical potential for both carriers\cite{ines_bis} [Eq. (\ref{mu})].
 This clarifies a confusing point in
the self-consistent treatment of interactions in higher-dimensional systems,
where the continuity of $V_{loc}$ is expected, even though not implemented
in the results. It would be interesting to define and then verify analogous
conditions to Eq. (\ref{conditions}).

The AC conductance matrix ${\bf G_3}(\omega )$ will be now expressed formally, where $\omega$ is the external frequency.
In the sequel, I switch to the Fourier transform  of expectation values in a time-dependent ground state. ${\bf G_3}%
(\omega )$ is a $3\times 3$ matrix with $g_{\alpha \beta }=\delta I_\alpha
/\delta V_\beta $ where $V_{1,2}=\mu _{1,2}/e$, $V_3=V_{gate}$, $I_{1,2}=\mp
j(\mp a,\omega )$ and $I_3$ the displacement current. 
A complete description with all the surrounding three-dimensional environment would be too
complex. Instead, assume that all the electric field lines emerging from the wire end
up on the gate, thus the latter carries an opposite charge to that on the
wire $Q(\omega )$. This ensures Kirchoff's law, $\sum_\alpha I_\alpha =0$
because, using the continuity equation, 
\begin{eqnarray}\label{kirchoff}
I_{3}(\omega)&=&-ie\omega Q(\omega)=\int_{-a}^a i\omega e\rho(x,\omega)\\\nonumber&=&-\int_{-a}^a \partial_x j(x,\omega)=-I_1(\omega)-I_2(\omega).
\end{eqnarray}
On the other hand, $V_{gate}(\omega )$ appears as a reference potential in Eqs. (\ref{mur},\ref{conditions}). Thus the
two constraints on the conductance matrix\cite{butticker_pretre} 
\begin{equation}
\label{constraints}\sum_\alpha g_{\alpha \beta }=0=\sum_\beta g_{\alpha
\beta }, 
\end{equation}
are ensured. Next focus on the first $2\times 2$ block of ${\bf G_3}%
(\omega )$ denoted ${\bf G_2}(\omega )$. Using Eqs. (\ref{conditions},\ref{jmu}), 
\begin{eqnarray}\label{I12}I_{1}=- j(- a,\omega )&=&\frac{e}h\left[\mu _{-}(-a,\omega)-\mu_{1}(\omega)\right],\nonumber \\I_2= j(+ a,\omega )&=&\frac{e}h\left[\mu _{+}(-a,\omega)-\mu_{2}(\omega)\right].\end{eqnarray} %
In order to express $I_\alpha $ to linear order in $\mu _{1,2}=\mu _{\pm
}(\mp a,\omega )$, it is sufficient to retain the linear dependence of $\mu
_{\pm }(\pm a,\omega )$, determined by some matrix ${\bf S}(\omega )$, 
\begin{equation}
\label{Somega}\left( 
\begin{array}{c}
\mu _{-}(-a,\omega ) \\ 
\mu _{+}(a,\omega ) 
\end{array}
\right) ={\bf S}(\omega )\left( 
\begin{array}{c}
\mu _{+}(-a,\omega ) \\ 
\mu _{-}(a,\omega ) 
\end{array}
\right) . 
\end{equation}
Combined with Eq. (\ref{I12}), this gives 
\begin{equation}
\label{two}{\bf G_2}(\omega )=\frac{e^2}h\left[ {\bf S}(\omega )-{\cal\bf I}%
\right] . 
\end{equation}
\label{S} In order to interpret ${\bf S}(\omega )$ as a ``scattering''
matrix, one can model the transition region by an N-channel system where electrons are
injected at the same quasi-equilibrium distribution. On the reservoir side
of $-a$, the total current is $J_{+}(\omega )-J_{-}(\omega )$, where $%
J_{+}(\omega )=Ne\mu _1(\omega )/h$ is the incident current, and $%
J_{-}(\omega )$ the unknown reflected current. The local continuity of the
current at $-a$ together with Eq. (\ref{jmu}) imply that $e\mu
_{-}(-a,\omega )/h=J_{-}(\omega )/N$ is the average reflected current per
channel. Similarly, $e\mu _{+}(a,\omega )/h$ is the average transmitted
current to the right reservoir. If the elements of ${\bf S}(\omega )$ are
denoted as follows 
\begin{equation}
{\bf S}(\omega )=\left( 
\begin{array}{cc}
R(\omega ) & T^{\prime }(\omega ) \\ 
T(\omega ) & R^{\prime }(\omega ) 
\end{array}
\right) , 
\end{equation}
then $T(\omega )$ [$R(\omega )$] can be viewed as the total dynamic transmission
[reflection] coefficient for the incident flux from the left to the right
reservoir [into the left reservoir]. $T^{\prime }$ and $R^{\prime }$ play
the same role for the right reservoir. Nevertheless, important
differences from the usual scattering approach should be stressed. The elements of ${\bf S}%
(\omega )$ determine directly the current or density, but are nonetheless
complex numbers. ${\bf S}(\omega )$ is not unitary, and in general not
symmetric unless there is a perfect reflection symmetry. In addition, $T(\omega
)+R(\omega )\neq 1$, and current conservation is ensured by the gate [Eq. (\ref
{kirchoff})]. Finally, using Eq. (\ref{two}), and letting the $\omega $
dependence be implicit :
\begin{equation}
\label{G3omega}{\bf G_3}=\frac{e^2}h\left( 
\begin{array}{ccc}
R-1 & T^{\prime } & 1-R-T^{\prime } \\ 
T & R^{\prime }-1 & 1-R^{\prime }-T \\ 
1-R-T & 1-R^{\prime }-T^{\prime } & R+R^{\prime }+T+T^{\prime }-2 
\end{array}
\right) 
\end{equation}
At zero frequency, $S(0)$ becomes real symmetric, and $T(0)+R(0)=T^{\prime
}(0)+R^{\prime }(0)=1$, but $R(0)$ can be negative. Then the DC conductance
is 
\begin{equation}
\label{gdc}g=g_{12}=-g_{{11}}=T(0)\frac{e^2}h. 
\end{equation}

As a first application of these boundary conditions, consider now a model where both total charges $Q_{\pm }=\int_{-a}^a\rho
_{\pm }(x)dx$ are conserved, 
\begin{equation}
\label{Qpm}[Q_{\pm },H]=0. 
\end{equation}
Then it is shown here that $T(0)=1$. For a quadratic Hamiltonian, this result was shown in Refs.[\onlinecite{ines,ines_bis}], and justifies the hypothesis adopted in Refs.[\onlinecite{DC}].

 In the Heisenberg representation, an operator $O$
evolves according to $i\hbar dO/dt=[H,O]+i\hbar \partial O/\partial t$, but $%
\partial O/\partial t=0$ in the stationary regime. For $r=\pm $, $\Phi _r$
is the canonical conjugate to $\rho _r$, thus $d\Phi _r/dt=-r\mu _r/\hbar $
[ Eq. (\ref{mur})]. Then $hd\rho _r/dt=r\partial _x\mu _r$ is an equation for
field operators that one can integrate between $-a$ and $a$ to get, using
Eq. (\ref{Qpm}), 
\begin{equation}
\label{muvariation}\mu _r(a,t)-\mu _r(-a,t)=rh\frac{dQ_r}{dt}=0. 
\end{equation}
On the other hand, the field $\mu _{+}(-a,t)$ cannot fluctuate but is equal
to $\mu _1$ [Eq. (\ref{conditions})], thus $\mu _{+}(a,t)=\mu _1$
at all times. Similarly, $\mu _{-}(-a,t)=\mu _{-}(a,t)=\mu _2$. Thus $T(0)=1$
and $R(0)=0$ [see Eq. (\ref{Somega})], and the DC
conductance is equal to $e^2/h$ [see Eq. (\ref{gdc}) or simply Eq. (\ref{jmu})].

A second application is to investigate ac transport in the simplest model (verifying Eq. (\ref{Qpm})) : the TLL model.
The matrix ${\bf S}(\omega )$ in Eq. (\ref{Somega}) can be computed in an
instructive way, through a ``transfer'' matrix ${\bf A}(\omega )$ such that 
\begin{equation}
\hbox{\boldmath${\mu}$}(a,\omega )={\bf A}(\omega )\hbox{\boldmath${\mu}$}%
(-a,\omega ),
\end{equation}
where {\boldmath${\mu}$} stands for the vector $(\mu_{+},\mu_{-})$. For this,
it is convenient to use the right- and left-propagating current modes $j_{\pm
}$, corresponding to up and down edge-excitations in a Hall bar,\cite
{wen_prl} that can be denoted ``quasiparticles''. The Hamiltonian density
is given by ${\mathcal{H}}(x)=h\left( j_{+}^2+j_{-}^2\right) /2e^2uK$, where $u$ and $K$
are interaction parameters.\cite{bosonisation} Without interactions, $j_{\pm
}=ev_F\rho _{\pm }$, $u=v_F$ and $K=1$. $j_{\pm }$ propagate freely at the
sound velocity $u$, thus 
\begin{equation}
\label{jaj-a}{\bf j}(a,\omega )=e^{i \hbox{\boldmath$\sigma_z$}\omega t_L}{\bf j}%
(-a,\omega ),
\end{equation}
where $t_L=2a/u$ is the transit time of the wire
and {\boldmath${\sigma_z}$} the z Pauli matrix. On the other hand, $j_{\pm }$ are related to $\rho
_{\pm }$ by simple diagonalization, but their relation to $\mu _{\pm }$ is of more use here :
\begin{eqnarray}\label{jMmu}{\bf j}(x,\omega)&=&\frac{e}h {\bf M}\hbox{\boldmath${\mu}$}(x,\omega)\nonumber\\ 
{\bf M}&=&\frac 1{1+\gamma }\left( 
\begin{array}{cc}
1 & -\gamma  \\ 
-\gamma  & 1
\end{array}
\right) ,\label{transfer}
\end{eqnarray}
where the coefficient $\gamma$ is given by\cite{ines} 
\begin{equation}
\label{gamma}\gamma =\frac{1-K}{1+K}.
\end{equation} Then ${\bf M}^{-1}$ can be obtained from ${\bf M}$ by $\gamma \rightarrow -\gamma 
$. 
Equations (\ref{jMmu},\ref{jaj-a}) yield the ``transfer'' matrix 
\begin{equation}
\label{transfertotal}{\bf A}(\omega )={\bf M}^{-1}e^{i \hbox{\boldmath${\sigma_z}$}\omega
t_L}{\bf M}.
\end{equation}
This allows to deduce the scattering matrix ${\bf S}(\omega )$ in Eq. (\ref
{Somega}), symmetric due to the reflection symmetry, 
\begin{eqnarray}
\label{Tomega}T(\omega )=T'(\omega)&=&(1-\gamma )\frac{e^{-i\omega t_L}+\gamma e^{i\omega
t_L}}{e^{-i\omega t_L}-\gamma ^2e^{i\omega t_L}},\\
\label{Romega}R(\omega )=R'(\omega)&=&\gamma \left[ 1-e^{i\omega t_L}T(\omega )\right] .
\end{eqnarray}
One can show that $|Det{\bf S}(\omega )|=1$, which is a constraint that has to hold
for any quadratic Hamiltonian with time-reversal symmetry. Note that
${\bf S}(\omega )$ depends solely on the intrinsic properties of the TLL
model; the boundary conditions (\ref{conditions}) allow to express the AC
conductance matrix through ${\bf S}(\omega )$, Eq. (\ref{G3omega}). I now analyze in more details the capacitive effects. The gate conductance is
\begin{equation}
\label{g33}g_{33}(\omega )=2\frac{e^2}h(1-\gamma )\frac{1-e^{i\omega t_L}}{%
1+\gamma e^{i\omega t_L}}.
\end{equation}
Thus the ``electro-chemical'' capacitance of the wire per unit length
with respect to the gate \cite{these_annales,blanter} 
$$
C=-\lim _{\omega \rightarrow 0}\left[ \frac{g_{33}(\omega )}{i\omega 2a}%
\right] =2\frac Ku\frac{e^2}h, 
$$
is proportional to the compressibility.\cite{bosonisation} This result can be
checked by minimizing the zero mode contribution to $H$ at fixed
total current, thus by minimizing $huQ^2/8aK+eQV_{gate}$. $C$ results from two capacitors in series: its value without interactions $C_0=e^2dn/dE=e^2 hv_F/2$, of purely kinetic origin, and the ``electrostatic'' capacitance $c$, obtained by evaluating Eq. (\ref{Vloc}),\cite{ines_bis,egger_charge,blanter} 
\begin{eqnarray}c=e\frac{\delta \rho }{\delta V_{loc}}&=&\frac {e^2}h\left(\frac u K-v_F\right)^{-1}.\\
\frac 1C&=&\frac 1c+\frac 1{C_0}. \label{series}
\end{eqnarray}
Interestingly, evaluating then differentiating Eqs. (\ref{Vloc},\ref{mu}) with respect to $\rho$ allows to recover Eq. (\ref{series}). In the TLL $\mu(x)=e^2\rho(x)/C+eV_{gate}$ justifying its interpretation as a local electrochemical potential for both carriers in Ref.[\onlinecite{ines_bis}]. But measuring $C$ gives the ratio $K/u$, leaving both $u$ and $K$ unknown. 
 For usual ballistic quantum wires, where $a$ is several $\mu m$, $2\pi/t_L\sim GHz$ is quite high. One is often in the regime $\omega t_L\ll 1$, where the non-dissipative part\cite{these_annales,blanter} of $g_{33}(\omega)$ [Eq. (\ref{g33})] is $-Im [g_{33}(\omega)]\simeq 2X-X^3(1-1/3K^2)/2$, with $X=C a\omega$. A strategy consists in measuring the leading term $2X$, then the subleading term that one divides by $X^3/2$ to infer $1-1/3K^2$, thus $K$. $u$ can be then determined from $C$. 

The underlying dynamics are now interpreted. Eq. (\ref{transfer}) is equivalent
to 
\begin{equation}
\label{scattering}\left( 
\begin{array}{c}
e\mu _{-}(x,\omega )/h \\ 
j_{+}(x,\omega )
\end{array}
\right) =\left( 
\begin{array}{cc}
\gamma  & 1+\gamma  \\ 
1-\gamma  & -\gamma 
\end{array}
\right) \left( 
\begin{array}{c}
e\mu _{+}(x,\omega )/h \\ 
j_{-}(x,\omega )
\end{array}
\right) ,
\end{equation}
so that the matrix on the right hand side can be viewed as a local
``scattering'' matrix.\cite{ines,ines_proc} Let us focus for instance on $x=-a$
where $\mu _{+}(-a,\omega )=\mu _1(\omega )$. When no charge is incident
from the left reservoir, i.e. $\mu _1=0$, then $j_{+}=-\gamma j_{-}$; $%
-\gamma $ is the reflection coefficient for a ``quasiparticle'' incident on the
contact. For repulsive interactions, $K<1$, thus $-\gamma <0$; a
``quasi-hole'' is reflected, in analogy with Andreev reflection.\cite{ines,ines_proc,these_annales,sandler} If no ``quasiparticle'' comes from
the right, i.e. $j_{-}=0$, then one finds
\begin{equation}
\label{muKmu}j_{+}=\frac eh(1-\gamma )\mu _1,
\end{equation}
and thus $K_a=2K/(1+K)=1-\gamma $ is the transmission coefficient for the
incident flux from the reservoir. $T(\omega )$ and $R(\omega )$ result from
the multiple reflections on the contacts, in analogy with a Fabry-Perot
resonator.\cite{ines,oreg} They have resonances at the collective modes of the finite
wire $\omega _n=uq$ for $q=2n\pi /2a$, at which $T=1$ and $R=0$. This is
because $e^{i\omega _n\hbox{\boldmath${\sigma_z}$}t_L}={\bf I}$, thus Eq. (\ref
{transfertotal}) becomes ${\bf A}(\omega _n)={\bf MM}^{-1}={\bf I}$.\cite
{these_annales} Note that since $j=j_{+}-j_{-}$, Eq. (\ref{muKmu})
obtained for $j_{-}=0$ yields the current at the interface of a
semi-infinite TLL and a Fermi liquid, the DC conductance becomes $%
g_a=K_ae^2/h$.\cite{ines,ines_proc,remark_chamon}

Indeed, all the above scattering matrices have been encountered in Refs.[\onlinecite{ines,ines_proc,these_annales}], where a TLL is connected perfectly to
noninteracting leads at $\pm a$.

 More generally, consider the case where $%
H_{int}$ is an arbitrary quadratic functional of $\rho $, vanishing for $%
|x|>a$. If an electron impinges at $t=0$ on $%
-a$, i.e. $\rho _{+}(x,t=0)=\delta (x+a)$, the transmitted [respectively reflected] charge to 
$a$ [at $-a$] at time $t$, i.e. $\rho_+(a,t)$ [$\rho_-(-a,t)$] is given by the function $M_{++}(a,-a,t)$
[$M_{-+}(-a,-a,t)$] whose Fourier transform {\em %
coincide} exactly with $T(\omega )$ (respectively $R(\omega )$) in Eq. (\ref
{scattering}). In addition, these functions determine the non local dynamic
conductivity at the contacts,\cite{ines,ines_proc} 
\begin{eqnarray}\label{sigmaTR}
\sigma (a,-a,\omega )&=&\frac{e^2}hT(\omega )\nonumber\\
\sigma (a,a,\omega )&=&\frac{e^2}h[1-R(\omega )].
\end{eqnarray}
The reservoirs can be modeled by an {\em %
external} potential that drops only at the contacts,\cite
{ines_proc,these_annales} so that, taking into account the constant gate
potential in $[-a,a]$, 
$$
E(x,\omega)=\left[ V_1-V_{gate}\right] \delta
(x+a)-\left[ V_2-V_{gate}\right] \delta (x-a). 
$$
Then $j(\pm a,\omega )=\sigma (a,\mp a,\omega
)\left[ V_{1,2}(\omega )-V_{gate}(\omega )\right] $, leading to the same ${\bf{G_3}(\omega)}$ [Eq. (\ref{G3omega})] by use of Eqs. (\ref{sigmaTR}%
,\ref{kirchoff}).\cite
{ines_proc,these_annales}

This is not a pure coincidence. The action being quadratic, the ground state properties are given exactly by minimizing it.  The equation of motion thus obtained imposes the continuity\cite{ines,ines_bis} of
both $j=e\left[ \mu _{+}-\mu _{-}\right] /h$ and that of Eq. (\ref{mu}), equals for instance $\mu(x)=hu\rho /K+eV_{gate}$ in the TLL. Here the $\omega$ dependence is implicit.
 Thus both $\mu _{+}$ and $\mu_{-}$ are continuous. For any $\left| x\right| >a$, $\mu _{\pm }(x)=hv_F\rho
_{\pm }(x)$ (see Eq. (\ref{mur})). The left reservoir injects electrons with
density $\rho _{+}(-a^{\left( -\right) },\omega )=\mu _1(\omega )/hv_F$ on
the noninteracting lead side, thus $\mu_{+}(-a^{\left( -\right) },\omega )=\mu_1(\omega)$. This 
fixes the continuous field $\mu_+$ on the interacting side, $\mu _{+}(-a^{\left( +\right)
},\omega )=\mu _1$, which is exactly Eq. (\ref{conditions}). Similar
reasoning holds symmetrically for the right reservoir. Also $\mu _{-}/h$ has
to be continuous at $-a$, thus it is equal to $v_F\rho _{-}$, the reflected
current. This is in accordance with the previous general interpretation of $%
{\bf S}(\omega )$ since one-dimensional leads model an effective channel.%
\cite{remark_chamon}

These arguments in favor of the noninteracting leads are restricted to a quadratic density Hamiltonian ${\mathcal{H}}(x)$ along the wire, although some of them might be extended if only $%
{\mathcal{H}}(\pm a)$ is quadratic. The role of backscattering in a TLL connected to
leads\cite{ines_proc,desordre,these_annales} was found to be controlled by the Fabry-Perot
dynamics recovered here by using Eq. (\ref{conditions}), but the
equivalence has to be checked and might be limited to linear transport. The
boundary conditions (\ref{conditions}), formulated without connecting
leads, are more general, and offer possibilities for future studies. One has
to reformulate the bosonisation procedure to compute the correlation functions. Implementing Eq. (\ref{conditions}) in a path integral formalism would give
access to the nonlinear regime, AC transport, and current fluctuations.
Conceptually, the scattering approach presented here can be extended to
situations where linear response theory fails, as will be the subject of a future study on edge
states and many channel systems.\cite{ines_hall}

 {\em Acknowledgments : } I
am indebted to T. Martin for reading the manuscript and for useful
criticism. I would like to acknowledge fruitful discussions with A.
Alekseev, Y. Blanter, D. Bernard, D. C. Glattli, T. Jolicoeur, and H. J.
Schulz.

{\em Note :} In a recent erratum,\cite{egger_err} Egger and Grabert modified
their boundary conditions for the TLL \cite{egger} by using self-consistent
arguments.\cite{egger_charge} Their corrected results agree with Ref.
[\onlinecite{ines}] and therefore with its present generalization.

\begin{thebibliography}{10}

\bibitem{scattering}
R. Landauer, Phil. Mag. {\bf 21}, 863 (1970); {Y. Imry}, {R. Landauer}, and {S.
  Pinhas}, Phys. Rev. B {\bf 31}, 6207 (1985).

\bibitem{butticker_pretre}
{M. B\"{u}ttiker}, {A. Pr\^{e}tre}, and {H. Thomas}, Phys. Rev. Lett. {\bf 70},
   4114  (1993).

\bibitem{AC_keldysh}
{N. S. Wingreen}, A.~P. Jauho, and Y. Meir, Phys. Rev. B {\bf 48}, R8487
  (1993-I); C. Bruder and H. Scholler, Phys. Rev. Lett. {\bf 72}, 1076 (1994).

\bibitem{ines}
I. Safi and H.~J. Schulz, Phys. Rev. B {\bf 52},  R17040  (1995).

\bibitem{maslov_pon}
D. Maslov and M. Stone, Phys. Rev. B {\bf 52}, R5539 (1995); V.~V. Ponomarenko,
  {\it ibid} R8666.

\bibitem{DC}
A.~Y. Alekseev, V. Cheianov, and J. Fr{\"o}hlich, Phys. Rev. B {\bf 54}, R17320
  (1996); A. Shimizu, J. Phys. Soc. Jpn. {\bf 65}, 1162 (1996); The first
  argument by Y. Oreg and A.~M. Finkel'stein, Phys. Rev. B {\bf 54}, R14265
  (1996) is similar,\cite{oreg} but is different from the second argument
  proposed also by A. Kawabata, J. Phys. Soc. Jpn. {\bf 65}, 30 (1996), and
  based on Kubo formula without reservoirs; this yields in general a different
  conductance \cite{ines_bis}.

\bibitem{ines_bis}
I. Safi, Phys. Rev. B {\bf 55},  R7331  (1997-II).

\bibitem{alekseev_goldstone}
A. Alekseev, V. Cheianov and J. Fr{\"{o}}hlich, cond-mat/9803346.

\bibitem{ines_proc}
I. Safi and H.~J. Schulz, in {\em Quantum Transport in Semiconductor Submicron
  Structures}, edited by B. Kramer (Kluwer Academic Press, Dordrecht, 1995); in
  {\em Correlated Fermions and Transport in Mesoscopic Systems}, edited by T.
  Martin, G. Montambaux, and J.~T. Van (Editions Frontieres, Gif-sur-Yvette,
  1996); Phys. Rev. B, {\bf 58}, August 1998 (in press), cond-mat/9803326.

\bibitem{these_annales}
In\`es Safi-Taktak, Ph.D. thesis, Laboratoire de Physique des Solides, 1996;
  Ann. Phys. (Paris) {\bf 22}, 463 (1997).

\bibitem{blanter}
Ya. M. Blanter, F. W. J. Hekking et M. B{\"u}ttiker, cond-mat/9710299.

\bibitem{AC_Kubo}
V.~V. Ponomarenko, Phys. Rev. B {\bf 52}, R8666 (1995); V.~A. Sablikov and
  B.~S. Shchamkhalova, JETP Lett {\bf 66}, 41 (1997).

\bibitem{kane_fisher_contacts}
C.~L. Kane and M.~P.~A. Fisher, Phys. Rev. B {\bf 52},  17393  (1995).

\bibitem{remark_chamon}
{C. de C. Chamon} and {E. Fradkin}, Phys. Rev. B {\bf 55}, 4534 ({1997-II}).

\bibitem{sandler}
N.~P. Sandler, C. de~C.~Chamon, and E. Fradkin, Phys. Rev. B {\bf 57},  12324
  (1998).

\bibitem{ines_hall}
In\`es Safi, in preparation.

\bibitem{bosonisation}
S. Tomonaga, Prog. Theor. Phys. {\bf 5}, 544 (1950); D.~C. Mattis and E.~H.
  Lieb, J. Math. Phys. {\bf 6}, 304 (1965).

\bibitem{remark_oreg}
For the specific case of a quadratic Hamiltonian, and in equilibrium, see also
  Ref.[\onlinecite{oreg}].

\bibitem{egger}
R. Egger and H. Grabert, Phys. Rev. Lett. {\bf 77},  538  (1996).

\bibitem{wen_prl}
X.~G. Wen, Phys. Rev. Lett. {\bf 64},  2206  (1990).

\bibitem{egger_charge}
R. Egger and H. Grabert, Phys. Rev. Lett. {\bf 79},  3463  (1997).

\bibitem{oreg}
Y. Oreg and A.~M. Finkel'stein, Phys. Rev. Lett. {\bf 74},  3668  (1993).

\bibitem{desordre}
A. Furusaki and N. Nagaosa, Phys. Rev. B {\bf 54}, R5239 (1996). See also D.
  Maslov, Phys. Rev. B {\bf 52}, R14368 (1995), but where results are
  restricted to the bulk.\cite{these_annales}.

\bibitem{egger_err}
R. Egger and H. Grabert, Phys. Rev. Lett. {\bf 80},  2255  (1998).

\end{thebibliography}

\end{document}